\documentclass[
    aps,prl,
    twocolumn,
    reprint,
    superscriptaddress,
    nofootinbib,
    floatfix,
    amssymb,
    long bibliography
]{revtex4-1}
\usepackage[utf8]{inputenc}
\usepackage{amsmath}
\usepackage{amsfonts}
\usepackage{graphicx}
\usepackage{bm}
\usepackage{dcolumn}
\usepackage{overpic}
\usepackage{mathtools}
\usepackage{dsfont}
\usepackage{nicefrac}
\usepackage[normalem]{ulem}
\usepackage{xcolor}
\definecolor{linkColor}{rgb}{0,0.3,0.7}
\usepackage{hyperref}
\hypersetup{colorlinks=true,
            allcolors=linkColor,
            pdfborder={0 0 0},
            pdfencoding = auto
            }

\definecolor{myGreen}{RGB}{19,132,23}

\begin{document}
\title{Local composition controls pattern formation in conserved active emulsions}

\author{Florian Raßhofer}
\affiliation{Arnold Sommerfeld Center for Theoretical Physics and Center for NanoScience, Department of Physics, Ludwig-Maximilians-Universit\"at M\"unchen, Theresienstra\ss e 37, D-80333 Munich, Germany}

\author{Erwin Frey}
\email[Corresponding author: ]{frey@lmu.de}
\affiliation{Arnold Sommerfeld Center for Theoretical Physics and Center for NanoScience, Department of Physics, Ludwig-Maximilians-Universit\"at M\"unchen, Theresienstra\ss e 37, D-80333 Munich, Germany}
\affiliation{Max Planck School Matter to Life, Hofgartenstraße 8, D-80539 Munich, Germany}

\date{\today}


\begin{abstract}
Phase separation in passive systems leads to uncontrolled droplet growth, limiting structural control in soft materials and cells.
We identify a generic mechanism to arrest coarsening based on chemical interconversion between molecular species with different diffusivities.
Sharp-interface theory and simulations show that when the faster-diffusing species becomes enriched inside droplets, composition gradients emerge that oppose mass influx.
This transport asymmetry stabilizes droplet sizes even without interaction asymmetries, offering a minimal route to regulate structure formation in active emulsions.
\end{abstract}

\maketitle
Phase separation is a fundamental organizing principle in soft-matter and biological systems, from synthetic emulsions\@~\cite{Lin.2023} and polymer blends\@~\cite{Knychala.2017}, to whole ecosystems\@~\cite{Siteur.2023}. 
It occurs when short-range interactions favor local (self-)association over mixing, causing multi-component systems to demix into coexisting phases.
In the absence of external driving and long-range interactions\@~\cite{Liu.1989,Glotzer.1994b,Muratov.2002,Kumar.2023,Winter.2025}, phase-separated domains inevitably coarsen: Larger domains grow at the expense of smaller ones through diffusive mass exchange. 
This unregulated growth, known as \textit{Ostwald ripening}\@~\cite{Ostwald.1897,Wagner.1961,Lifshitz.1961}, is driven by interfacial tension and constrains structural control in both materials and living systems.

This limitation has received renewed attention with the discovery of biomolecular condensates\@~\cite{Hyman.2014,Banani.2017}.
These membraneless organelles participate in essential biological processes, including midcell localization\@~\cite{Schumacher.2017,Bergeler.2018}, carbon fixation\@~\cite{Wunder.2018,He.2023}, ribosome assembly\@~\cite{Feric.2016}, and the pathology of several diseases\@~\cite{Alberti.2019,Alberti.2021}.
While they are believed to form via liquid–liquid phase separation\@~\cite{Brangwynne.2009,Boeynaems.2018,Fritsch.2021}, the spontaneous emergence and unregulated growth predicted by passive phase separation are difficult to reconcile with the precise spatial and temporal control required in cells.
This suggests that living systems employ active mechanisms to regulate condensate formation and morphology\@~\cite{Snead.2019,Soeding.2020}. 

One class of mechanisms involves chemical reactions that modify phase behavior\@~\cite{Glotzer.1994,Glotzer.1995,Weber.2019,Li.2020,Jülicher.2023}, giving rise to stable droplet sizes\@~\cite{Carati.1997,Zwicker.2015,Wurtz.2018,Cates.2010,Osmanovic.2023,Caballero.2023,Bauermann.2025}, droplet self-propulsion\@~\cite{Demarchi.2023,Haefner.2024,Goychuk.2024,Jambon-Puillet.2024}, division\@~\cite{Zwicker.2017}, and persistent capillary waves\@~\cite{Rasshofer.2025}.
Previous studies have mostly examined reactions that modulate component interactions\@~\cite{Zwicker.2015,Wurtz.2018}, e.g., through post-translational modifications\@~\cite{Rai.2018,Hofweber.2019,Owen.2019}.
Here, we introduce a distinct form of active regulation: 
\emph{Interconversion between molecular states of different diffusivity.}

Such conversion is known to drive pattern formation in dilute (mass-conserving) reaction–diffusion [(Mc)RD] systems\@~\cite{Turing.1952,Halatek.2018b,Brauns.2020,Frey.2022}, and has been linked to the emergence of thermophoretic fluxes\@~\cite{Liang.2022}.
However, its consequences for dense, interacting mixtures remain largely unexplored\@~\cite{Aslyamov.2023,Menou.2023,Osmanovic.2025}.
A key open question is whether conversion between states of unequal diffusivity can arrest coarsening in systems with strong interactions. 
To address this, we study an incompressible ternary mixture of two solutes, A and\@~B, and a solvent\@~S; a generalization to arbitrary numbers of components is discussed in Ref.\@~\cite{pre}. 
The solutes are assumed thermodynamically equivalent---sharing identical interactions and internal energies---but differ in diffusivity, and interconvert through chemical reactions.
This minimal model extends classical two-component McRD systems\@~\cite{Halatek.2018b,Frey.2022} to the regime of strong interactions, enabling us to disentangle the roles of short-range interactions and mobility regulation.

Combining finite-element (FEM) simulations, linear stability analysis, and sharp-interface theory, we identify two distinct pattern-forming regimes:
When pairwise interactions are too weak to induce phase separation, patterns emerge solely via a generalized \textit{mass-redistribution instability}\@~\cite{Brauns.2020,Frey.2022} and ultimately coarsen into a single dense domain.
In contrast, when interactions favor phase separation, coarsening can be arrested if chemical activity locally enriches the faster diffusing species within high-density domains.
This identifies a previously unrecognized mechanism by which chemical activity can arrest coarsening solely through driven transport asymmetries. 
Detailed calculations and further results on phase-coexistence outside the arrested regime are presented in Ref.\@~\cite{pre}.

\textit{Model.}
In thermodynamic equilibrium, the statistical properties of the mixture are described by a Flory-Huggins free energy $\mathcal{F}$\@~\cite{Rubinstein.2003,Huggins.1941,Flory.1942}, which depends on the local volume fractions $\phi_{\mathrm{A}/\mathrm{B}}$.
Introducing the conserved \emph{solute volume fraction} ${\rho=\phi_\mathrm{A}+\phi_\mathrm{B}}$, and the \textit{local composition} ${s=(\phi_\mathrm{A}-\phi_\mathrm{B})/(\phi_\mathrm{A}+\phi_\mathrm{B})}$, one finds\@~\cite{pre}
\begin{align}
    \mathcal{F} [\rho,s]
    = 
    \frac{k_BT}{\nu}  \int \hspace{-1mm} \mathrm{d}\boldsymbol{r} 
    \bigg[ 
    \frac{\tilde\kappa}{2} \left(\boldsymbol{\nabla}\rho\right)^2+f(\rho) + \rho \, g(s) 
    \bigg]
    \, ,
    \label{eq:FreeEnergy}
\end{align}
where we assume that all components occupy the same molecular volume\@~$\nu$. 
For brevity, we occasionally refer to\@~$\rho$ as the solute density, although it denotes a volume fraction.
The first two terms in Eq.\@~\eqref{eq:FreeEnergy} describe a symmetric binary mixture:
Density gradients are penalized by a stiffness coefficient\@~$\tilde\kappa$, and the local free energy density ${f(\rho) = \rho \log(\rho) + (1-\rho) \log (1-\rho) + \chi\rho (1-\rho)}$ includes entropic and energetic contributions, with $\chi$ denoting the Flory–Huggins parameter\@~\cite{Rubinstein.2003}.
For ${\chi > 2}$, $f(\rho)$ becomes non-convex, and the system separates into coexisting phases characterized by the local minima\@~$\rho_\pm$ of\@~$f(\rho)$.
The last term, ${\rho \, g(s)}$, describes the entropy of mixing between A and B particles.
The function ${g(s)=\frac{1+s}{2}\log\left(\frac{1+s}{2}\right)+\frac{1-s}{2}\log\left(\frac{1-s}{2}\right)}$ is strictly convex, ensuring thermodynamic stability against compositional fluctuations. 

In the absence of chemical reactions, relaxation towards thermodynamic equilibrium follows the principles of non-equilibrium thermodynamics\@~\cite{Groot.1984,Onsager.1930,*Onsager.1931}. 
Each species ${i\in\{\mathrm{A,B}\}}$ satisfies a continuity equation with diffusive flux ${\boldsymbol{J}_{i}= - M_{i}(\{\phi_j\}) \boldsymbol{\nabla}\bar \mu_i}$, where ${\bar \mu_{i}=\nu \, \delta\mathcal{F}/\delta \phi_{i}}$ denote the exchange chemical potentials.
Neglecting cross-diffusion\@~\cite{pre}, we choose effective mobilites ${M_i = D_i \phi_i (1 - \rho)}$, to recover Fick's law with constant\@~$D_i$ in the dilute, non-interacting limit.

In addition, we include local interconversion ${\mathrm{B} \rightleftarrows \mathrm{A}}$ with rates ${k_{\to}}$ and ${k_{\leftarrow}}$, giving rise to a reactive flux ${\tilde{\mathcal{R}} = k_{\to} \phi_\mathrm{B} - k_{\leftarrow} \phi_\mathrm{A}}$.
Unless stated otherwise, we choose ${k_{\to} = \rho/\tau}$ and ${k_{\leftarrow} = \rho_0/\tau}$, where\@~${\tau}$ sets the reaction timescale, and ${\rho_0=0.5}$ is fixed throughout the manuscript.
This yields a unique steady-state composition ${\bar s(\rho)=(\rho-\rho_0)/(\rho+\rho_0)}$, implying enrichment of A in dense regions (${\partial_\rho \bar s>0}$), breaking the inherent ${\mathrm{A}\leftrightarrow\mathrm{B}}$ symmetry of the free energy\@~[Eq.\@~\eqref{eq:FreeEnergy}].
Such density-enhanced ${\mathrm{B}\to\mathrm{A}}$ conversion may arise from crowding-induced conformational switching\@~\cite{Dong.2010,Wen.2021}, phase-specific catalysis\@~\cite{Harris.2025,O'Flynn.2021}, or active mobility regulation\@~\cite{Liu.2013,Fu.2012}. 
Our results, however, do not rely on this specific parametrization: 
Any rate law that generates a unique, stable, and monotonic fixed point\@~${\bar s(\rho)}$ produces the same qualitative behavior. 

To identify the relevant control parameters, we nondimensionalize space and time as ${\boldsymbol{r} \to a\, \boldsymbol{r}}$ and ${t\rightarrow (a^2/\bar D) \, t}$, using the particle diameter ${a=\nu^{1/d}}$ and the mean diffusivity ${\bar D = (D_\mathrm{A}+D_\mathrm{B})/2}$ as natural units. 
The resulting dimensionless equations read\@~\cite{pre}
\begin{subequations}
\label{eq:Dynamics}
\begin{align} 
	\partial_t \rho 
	=&\boldsymbol\nabla \left\{M(\rho)(1+D s) \boldsymbol{\nabla}\left[\mu+\log(1+Ds)\right] \right\}\, ,\label{eq:rhoEvolution}\\
	\partial_t [s\rho]  
	=& \boldsymbol\nabla \big\{M(\rho)\big[(D+s) \boldsymbol{\nabla}\mu +\boldsymbol{\nabla}s\big]\big\} + \dfrac{{\cal R}(\rho,s)}{\ell^2}\,,\label{eq:sEvolution}
\end{align}
\end{subequations}
with ${M(\rho)=\rho \, (1-\rho)}$, the binary chemical potential ${\mu=-\kappa\,\boldsymbol{\nabla}^2\rho+f^\prime(\rho)}$ (${\kappa=\tilde\kappa \, a^{-2}}$), and  ${\mathcal{R} = \tau \, \tilde{\mathcal{R}}}$.
Two dimensionless parameters capture the departure from ordinary mixture dynamics:
The diffusivity contrast ${D=(D_\mathrm{A}-D_\mathrm{B})/(D_\mathrm{A}+D_\mathrm{B})}$ encodes how transport asymmetry biases mass-fluxes away from the case of a binary mixture, which is recovered for ${D=0}$.
The parameter ${\ell^2=\tau\bar D/(2a^2)}$ compares the typical diffusion length during a reaction time, ${l_\tau^2\sim\tau\bar D}$, to the microscopic particle size\@~$a$, setting the characteristic (dimensionless) length scale over which the composition field responds to local density variations.

\textit{Limiting behavior.}
In the absence of chemical activity, or when reactions satisfy detailed balance\@~\cite{pre}, the steady state is characterized by uniform composition (${\boldsymbol{\nabla}s=0}$) and chemical potential (${\boldsymbol{\nabla}\mu=0}$).
Consequently, phase coexistence is governed by the equilibrium conditions of a binary mixture.

In the fast-reaction limit\@~(${\ell \ll 1}$), local composition relaxes instantaneously to its reactive steady state, ${s = \bar s(\rho)}$, allowing for adiabatic elimination.
In this case, $\rho$ obeys conserved gradient dynamics ${\partial_t \rho = \boldsymbol{\nabla} \left\{ M(\rho)[1+D\bar s(\rho)]\, \boldsymbol{\nabla} \mu_\text{eff} \right\}}$, where the effective chemical potential ${\mu_\text{eff}=\delta\mathcal{F}_\text{eff}/\delta\rho}$ follows from the free energy functional~\cite{pre}
\begin{align}
    \mathcal{F}_\text{eff}[\rho]
    \!=\!
    \int \hspace{-0.5mm}\mathrm{d}\boldsymbol{r} \, \Big\{\frac{\kappa}{2}(\boldsymbol{\nabla}\rho)^2 \!+\! f(\rho) \!+\! \int^\rho  \hspace{-0.5mm}\log[1\!+\!D \, \bar s(\tilde\rho)]\Big\}.
\end{align}
Despite being out of equilibrium, the dynamics map to an effective binary mixture whose chemical potential is corrected by a term ${\log[1+D \,\bar s(\rho)]}$.
Hence, genuinely non-equilibrium behavior arises only for intermediate activity levels.

\begin{figure}
    \centering
    \begin{overpic}[]{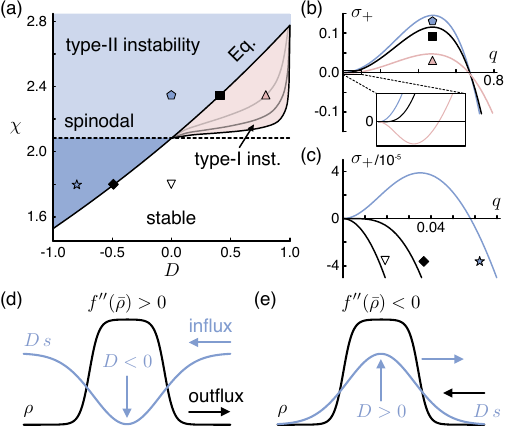}
            \put(48.5,77){\rotatebox{45.5}{\normalsize \sffamily \eqref{eq:massRedistributionInstability}}}
    \end{overpic}
    \caption{\textbf{Linear stability analysis}. 
    The homogeneous steady state is stable (white) or displays a \mbox{type-I} (red/light gray) or \mbox{type-II} (blue/gray) instability.
    (a) Stability diagram as a function of $D$ and $\chi$ for fixed ${\bar\rho=0.4}$, ${d=2}$, and ${\ell\in\{5,10,20\}}$ (light gray, gray, and black lines). 
    In the presence of driven chemical reactions, the boundary of the type-II regime (black line) is determined by Eq.\@~\eqref{eq:massRedistributionInstability}. 
    In the equilibrium limit, a homogeneous state is linearly unstable above the (dashed) spinodal line ${\chi=1/[2\bar{\rho}(1 - \bar{\rho})]}$.
    (b,\@~c) Upper branch of the dispersion relation, $\sigma_+(q)$, for parameters corresponding to the different markers in subpanel (a) and ${\ell=20}$.
    (d,\@~e) Composition weighted diffusivity $D\, s$ (blue/gray) for a weakly perturbed homogeneous density\@~$\rho$ (black). 
    Horizontal arrows indicate composition- (blue/gray) and chemical-potential-driven (black) fluxes, respectively.
    }
    \label{fig:LSAFigure}
\end{figure}

\textit{Stability.}  
Linear stability analysis of the homogeneous steady state\@~${(\bar\rho,\bar s)}$ yields a dispersion relation with two branches, $\sigma_\pm(q)$, which satisfy ${ \sigma_+ \geq \sigma_-,\, \forall q}$, and are real-valued over nearly the entire parameter range; complex eigenvalues only occur in a narrow, linearly stable regime~\cite{pre}.
As detailed in Fig.\@~\ref{fig:LSAFigure}(a), a long-wavelength (type-II) pattern-forming instability\@~\cite{Cross.1993} occurs when~\cite{pre}
\begin{align}
    \big[
    1 + D \, \bar s (\bar \rho)
    \big] \, 
    f^{\prime \prime} (\bar \rho) 
    < -D\, 
    \partial_\rho {\bar s} (\bar\rho)
    \,,
\label{eq:massRedistributionInstability}
\end{align}
generalizing the mass-redistribution instability known from two-component McRD systems\@~\cite{Halatek.2018b,Frey.2022}.
Without chemical activity or for equal diffusivities\@~(${D = 0}$), the condition reduces to the classical spinodal criterion ${f^{\prime \prime} (\bar \rho) < 0}$ for phase separation in a binary mixture\@~\cite{Bray.2002,Rubinstein.2003}.

The instability condition, Eq.\@~\eqref{eq:massRedistributionInstability}, reflects a competition between composition-driven and chemical-potential-driven fluxes\@~[Fig.\@~\ref{fig:LSAFigure}(d,e)]:
To illustrate this, consider a small local density increase above the homogeneous state.
This shifts local chemical equilibrium, inducing a change in composition, which has the same sign as the density change\@~[${\partial_\rho {\bar s} > 0}$, Fig.\@~\ref{fig:LSAFigure}(e)].
The resulting composition gradient drives a mass flux ${-DM(\rho)\boldsymbol{\nabla} s \approx - D M(\bar\rho)\partial_\rho {\bar s}\boldsymbol{\nabla}\rho}$\@~[Eq.\@~\eqref{eq:rhoEvolution}].
If A diffuses more slowly (${D<0}$), the flux points up the density gradient and amplifies the perturbation\@~[Fig.\@~\ref{fig:LSAFigure}(d)].
In contrast, outside the spinodal regime [${f^{\prime \prime} (\bar \rho)>0}$], chemical potential gradients drive a restoring flux with a positive effective diffusion constant ${M(1 + D \bar{s}) f^{\prime \prime} (\bar \rho)}$\@~[Eq.\@~\eqref{eq:rhoEvolution}].
Instability sets in when the composition-driven influx dominates, i.e., when Eq.\@~\eqref{eq:massRedistributionInstability} holds.
Conversely, if the faster-diffusing species accumulates in dense regions [${D\, \partial_\rho {\bar s} > 0}$,\@~Fig.\@~\ref{fig:LSAFigure}(e)], composition-driven fluxes oppose mass accumulation, and a \mbox{type-II} instability is only possible within the spinodal regime\@~[${f^{\prime\prime}(\bar\rho)<0}$].

When the mass-redistribution criterion\@~[Eq.\@~\eqref{eq:massRedistributionInstability}] is not satisfied, long-wavelength modes remain stable.
Simultaneously, interfacial stiffness $\kappa$ suppresses short-wavelength fluctuations, leaving only an intermediate wavenumber interval in which modes can grow for sufficiently strong interactions\@~\mbox{[large $\chi$, Fig.\@~\ref{fig:LSAFigure}(a, b)]}.
Unlike the \mbox{type-II} criterion\@~[Eq.\@~\eqref{eq:massRedistributionInstability}], which is independent of the diffusive length scale\@~$\ell$, the onset of this \mbox{type-I} instability depends on\@~$\ell$\@~[Fig.\@~\ref{fig:LSAFigure}(a)].
This reflects their different physical origins: 
\mbox{Type-II} instabilities are purely diffusive, governed by $\mathcal{O}(q^2)$ terms in the dispersion relation, whereas \mbox{type-I} instabilities require a sign change in the higher-order $\mathcal{O}(q^4)$ term, which depends on $\ell$\@~\cite{pre}.

\begin{figure}
    \centering
    \includegraphics[]{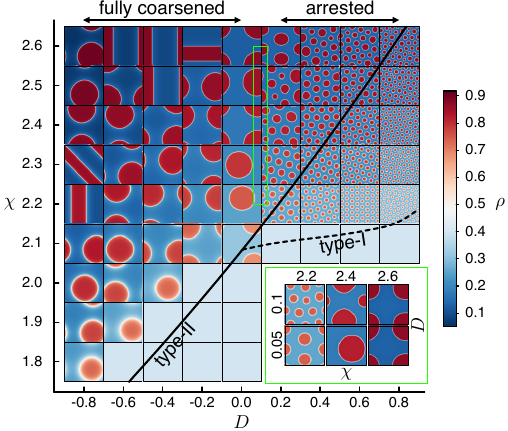}
    \caption{\textbf{Phase diagram}. Snapshots of the conserved density\@~$\rho$ at time ${t=10^8/(1+D)}$, obtained from FEM simulations in a square domain of size ${L=200}$ with periodic boundary conditions.
    The system is initialized with small perturbations around the homogeneous steady state.
    Color indicates local density (colorbar); in grayscale renderings, interfaces appear white.
    The inset shows snapshots for ${D\leq0.1}$ (green box).
    Dashed and solid lines mark the boundaries of the type-I and type-II unstable regimes, respectively (Fig.\@~\ref{fig:LSAFigure}). 
    All simulations use ${\ell=20}$ and the other parameters as in\@~Fig.\@~\ref{fig:LSAFigure}.
    }
    \label{fig:phaseDiagram}
\end{figure}

\textit{Phase diagram.} 
To explore the nonlinear dynamics, we performed FEM simulations over a broad parameter range\@~\cite{SI}. 
Within the linearly unstable region, the outcome depends sensitively on the diffusivity contrast~(Fig.\@~\ref{fig:phaseDiagram}): 
For ${D \le 0}$, the system invariably coarsens into macroscopic phase separation, whereas for ${D > 0}$ complete phase separation occurs only in a narrow region ${D \lesssim 0.1}$; at larger\@$D$ the system settles into an \emph{ensemble of coexisting finite-size droplets}. 

While arrested coarsening is typically expected near the onset of a \mbox{type-I} instability\@~\cite{Cross.2009}, we observe that enrichment of the faster-diffusing species in dense regions robustly arrests coarsening even within the type-II unstable regime, provided phase separation is driven by pairwise interactions\@~(${\chi>2}$). 
In contrast, when pattern formation is driven by composition gradients and opposed by chemical-potential fluxes, no stable finite-size structures emerge:
As shown in Fig.\@~\ref{fig:phaseDiagram}, transient patterns can still emerge in the thermodynamically stable regime\@~(${\chi \leq 2}$) via the generalized mass-redistribution instability [Eq.\@~\eqref{eq:massRedistributionInstability}].
However, as for dilute two-component McRD systems\@~\cite{Brauns.2021,Weyer.2023}, these patterns invariably coarsen, regardless of the specific reaction kinetics $\mathcal R$\@~\cite{SI}.
This shows that coarsening arrest requires attractive short-range interactions combined with opposing fluxes from active composition gradients---not the reverse.

\textit{Arrested coarsening.} 
To uncover the mechanism of arrested coarsening, we consider a dilute solution where droplets are well separated and interact only through the common far-field density\@~$\rho_\infty$.
In this limit, the dynamics reduce to those of an isolated spherical droplet in a homogeneous background.
Coarsening is arrested if Eq.\@~\eqref{eq:Dynamics} admits a stable steady-state solution at finite droplet radius\@~$R$.
If no such fixed point exists, droplets dissolve\@~(${R \to 0}$) or grow without bound\@~(${R \to \infty}$).

The local solute flux ${\boldsymbol{J}_\rho = -M(\rho)(1+Ds)\,\boldsymbol{\nabla}\eta}$ [Eq.\@~\eqref{eq:rhoEvolution}] is governed by the mass-redistribution potential ${\eta (\boldsymbol{r}) \equiv \mu(\boldsymbol{r}) + \log [1+D\,s(\boldsymbol{r})]}$.
In steady state, vanishing mass flux requires ${\boldsymbol{\nabla}\eta=0}$.
Evaluating this condition at the droplet interface and in the far-field yields 
\begin{align}
    \mu_\infty -\mu(R)  
    =     
    \log\left[1+D s(R)\right] - \log\left[1+D s_\infty\right]  
    \, ,
\label{eq:NoFluxCondition}
\end{align}
where ${\mu_\infty = f^\prime(\rho_\infty)}$ and ${s_\infty = \bar s(\rho_\infty)}$ are set by local thermodynamic and chemical equilibrium.
Equation\@~\eqref{eq:NoFluxCondition} shows that a finite droplet size can be maintained only if the chemical potential difference is exactly balanced by a change in local composition.
Since\@~$\mu(R)$ and\@~$s(R)$ depend on the stationary profiles, determining this balance requires the full solution for $\rho(r)$ and $s(r)$.
\begin{figure}
    \centering
    \includegraphics[]{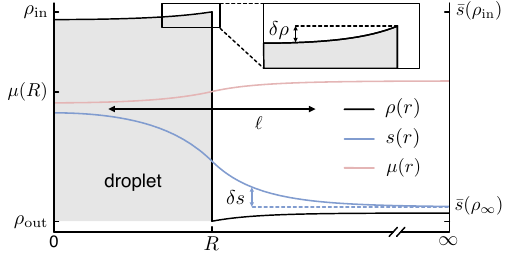}
    \caption{
    \textbf{Sharp-interface limit} for an isolated droplet. An infinitely sharp interface at ${r=R}$ separates the high-density interior (gray) from the low-density exterior (white). Within each domain, the solute density\@~$\rho$ (black) and chemical potential\@~$\mu$ (red/light gray) remain close to their equilibrium values\@~$\rho_\text{in/out}$ and $\mu(R)$. 
    The composition\@~$s$\@~(blue) varies on a characteristic length scale\@~$\ell$, and relaxes towards chemical equilibrium in the far field\@~[${s \to \bar s(\rho_\infty)}$].
    }
    \label{fig:sharpInterface}
\end{figure}

\emph{Sharp interface limit}. 
Steady-state solutions of Eq.\@~\eqref{eq:Dynamics} can be obtained in the asymptotic regime (${w \ll R, \ell}$), where the interfacial width\@~$w$ is negligible compared with the droplet radius $R$ and the diffusive length scale $\ell$ (Fig.\@~\ref{fig:sharpInterface}).
This scale separation allows for several simplifications:

(i) 
Across the narrow interface, the composition is nearly uniform\@~(${\boldsymbol{\nabla} s\approx 0}$).
Stationarity (${\boldsymbol{\nabla}\eta=0}$), then, implies ${\boldsymbol{\nabla}\mu\approx 0}$, i.e., local thermodynamic equilibrium.
Consequently, the interfacial chemical potential $\mu(R)$ and the coexisting densities\@~$\rho_\text{in/out}$ adjacent to the interface\@~(Fig.\@~\ref{fig:sharpInterface}) are determined by a standard common-tangent construction\@~\cite{pre,Bray.2002}.

(ii)
The droplet interior and exterior can be treated as quasi-homogeneous compartments, with only small density variations\@~(Fig.\@~\ref{fig:sharpInterface}).
Accordingly, we linearize the steady-state equations\@~[Eq.\@~\eqref{eq:Dynamics}] around the equilibrium coexistence densities, ${\rho = \rho_\text{in/out} + \delta \rho}$, imposing ${\delta \rho(R) = 0}$ at the interface.
Linearization of the composition field\@~$s$ is more subtle:
For large droplets (${R\gg \ell}$), $s$ varies between the two reactive equilibria at the droplet center\@~[${s\approx \bar s(\rho_\text{in})}$] and in the far-field\@~[${s=\bar s(\rho_\infty)}$].
Thus, compositional variations\@~${\delta s\approx\mathcal{O}[\bar s(\rho_\text{in})-\bar s(\rho_\infty)]}$ are generally not small.
In contrast, for small droplets (${R\ll \ell}$), the composition inside the droplet cannot fully relax to chemical equilibrium and instead remains nearly uniform throughout the system; see Fig.\@~\ref{fig:sharpInterface}.
In this limit, we write ${s\approx s_\infty+\delta s}$ and expand Eq.\@~\eqref{eq:Dynamics} to leading order in ${\delta s}$.
Detailed validity bounds and alternative linearization schemes are discuss in Ref.\@~\cite{pre}. 

(iii) Finally, we use the stationarity condition, ${\eta(\delta\rho,\delta s) = \mu_\infty + \log(1 + D s_\infty)}$, which links compositional and density variations, to eliminate\@~$\delta \rho$. 
To leading order in\@~$\delta s$, the steady-state equations [Eq.\@~\eqref{eq:Dynamics}] then reduce to a set of piecewise-linear differential equations
\begin{align}
    \hspace{-1mm}
    K_\alpha \boldsymbol{\nabla}^2 \, \delta s_\alpha(r)
    + \frac{K_\alpha}{\ell_\alpha^2}\,
    \big[
    \delta s_\alpha^0-\delta s_\alpha(r)
    \big]
    = 0 \, ,
    \label{eq:HelmholtzEquation}
\end{align}
where $K_\alpha$ and $\ell_\alpha$ with ${\alpha \in \{\text{in, out}\}}$ denote the region-specific mobility and screening length. 
While the outer offset ${\delta s_\text{out}^0=0}$ vanishes (${s\to s_\infty}$ as ${r\to\infty}$), the inner offset\@~$\delta s_\text{in}^0$ characterizes the composition shift at the center of a large droplet (${R \gg \ell_\text{in}}$)\@~\cite{pre}.

Solutions to Eq.\@~\eqref{eq:HelmholtzEquation}, are obtained by enforcing regularity at the origin (${\partial_r s|_{r=0}=0}$), and matching the density to its far-field value [${\lim_{r\to\infty}\rho(r)=\rho_\infty}$]. 
Integration across the interface further imposes flux continuity: ${K_\text{in}\partial_r \delta s_\text{in}(R)=K_\text{out}\partial_r \delta s_\text{out}(R)}$.
This yields a nonlinear self-consistency condition for the stationary radius\@~$R$, which is solved numerically\@~\cite{SI}.

\begin{figure}
    \centering
    \includegraphics[]{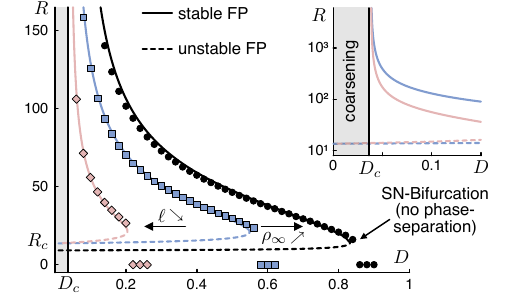}
    \caption{\textbf{Single droplet steady states} in a supersaturated solution. Stable (solid) and unstable (dashed) stationary radii obtained from sharp-interface theory and FEM simulations (bullets) in a three-dimensional\@~(${d=3}$) spherically symmetric system (${L=15\,\ell}$)\@~\cite{SI} for ${\chi=2.4}$ (${\rho_-=0.17}$), and ${\{\rho_\infty,\ell\}=}$ ${\{0.185,200\}}$ (black), ${\{0.18,200\}}$ (blue/gray), ${\{0.18,100\}}$ (red/light gray). The inset shows the asymptotic behavior near the onset $D_c$ of arrested coarsening\@~[Eq.\@~\eqref{eq:criticalD}].}
    \label{fig:singleDroplet}
\end{figure}

Explicit calculation steps and analytical results are presented in Ref.\@~\cite{pre}.
Depending on the far-field density $\rho_\infty$, and the diffusivity contrast\@~$D$, one finds either one, two, or no steady-state solutions\@~(Fig.\@~\ref{fig:singleDroplet}).
Here, we focus on the supersaturated case (${\rho_\infty>\rho_-}$), for which droplet growth occurs in an equilibrium mixture\@~\cite{Bray.2002}; the undersaturated scenario is discussed in Ref.\@~\cite{pre}.

In the equilibrium limit ($D=0$), the system admits a single stationary solution\@~(Fig.\@~\ref{fig:singleDroplet}) at the critical nucleation radius $R_c$ of a binary mixture, which corresponds to an unstable fixed point\@~\cite{Bray.2002}.
For increasing diffusivity contrast\@~$D$, this solution persists with little change, showing that the nucleation barrier is only weakly affected by the non-equilibrium drive.
Beyond a critical threshold $D_c$, a second stationary solution emerges\@~(Fig.\@~\ref{fig:singleDroplet}), corresponding to a stable droplet of finite size.
The onset of this arrested regime can be estimated as\@~\cite{pre}
\begin{align}
    D_c= \frac{f''(\rho_-)(\rho_\infty-\rho_-)\left(1+\sqrt{\frac{\partial_s\mathcal R|_\text{out}}{\partial_s \mathcal R|_\text{in}}}\,\right)}{\left(\partial_s \log\mathcal R|_\text{out}\right)^{-1} - \left(\partial_s \log\mathcal R|_\text{in}\right)^{-1}} \, ,
    \label{eq:criticalD}
\end{align}
where $|_\text{in/out}$ denotes evaluation at $(\rho_\pm,s_\infty)$.
Equation\@~\eqref{eq:criticalD} shows that stronger supersaturation requires a larger diffusivity contrast to arrest coarsening, while the denominator quantifies the asymmetry of the reaction’s sensitivity to composition changes inside versus outside the droplet, with greater asymmetry lowering the threshold.
At\@~${D=D_\text{SN}}$, which increases with\@~$\rho_\infty$ and decreases with\@~$\ell$, both fixed points vanish in a saddle–node\@~(SN) bifurcation, signaling complete suppression of phase separation\@~(Fig.\@~\ref{fig:singleDroplet}). 

Comparison with FEM simulations of a three-dimensional spherical droplet\@~\cite{SI} shows quantitative agreement for both the steady droplet radius and the onset of arrested coarsening\@~(Fig.\@~\ref{fig:singleDroplet}).
Only for stronger supersaturation and near the saddle-node bifurcation, the sharp-interface theory weakly deviates from the FEM results.
These discrepancies correlate with larger $D$ and $\rho_\infty$, which drive the density profile away from the equilibrium plateau values\@~$\rho_\text{in/out}$, indicating the gradual breakdown of our linearization scheme.

\textit{Conclusions}.
We have shown that interconversion between molecular states of different diffusivity---but identical interactions---can arrest coarsening when the faster species is enriched in dense regions.
Crucially, the resulting droplet size is not set by the wavelength of a pattern-forming instability, but emerges from a balance of mass fluxes driven by chemical potential and composition gradients\@~[Eq.\@~\eqref{eq:NoFluxCondition}]. 

This mechanism differs from established models of coarsening arrest in dense interacting mixtures, which rely on reactions converting between strongly interacting and freely diffusing species\@~\cite{Donau.2023,Wurtz.2018,Zwicker.2015}.
In practice, chemical modifications of the solutes likely affect both mobility and interactions, making it challenging to isolate one effect from the other.
However, mobility regulation may be particularly relevant for membrane-associated phase separation\@~\cite{Su.2016,Banjade.2014}, where protein mobility depends on the number of anchors\@~\cite{Khmelinskaia.2016,Khmelinskaia.2018}, lipid-binding domains\@~\cite{Knight.2010,Ziemba.2013}, and membrane inclusion size\@~\cite{Saffman.1975,Javanainen.2017}. 

These properties could be dynamically tuned through reversible modifications, such as lipidation\@~\cite{Jiang.2018}, enabling changes in mobility without substantially altering inter-particle interactions.
Similar effects may also arise in the cytoplasm: Phosphorylation-regulated binding to slow-moving structures, such as RNA clusters\@~\cite{Griffin.2011}, could transiently reduce molecular mobility in the bulk, offering an alternative route to mobility-dependent regulation.

From a broader perspective, our results uncover a general mechanism for regulating mesoscale organization in active mixtures: 
Positive feedback between local density and effective mobility can counteract coarsening and stabilize finite-size domains.
An analogous feedback mechanism operates in ecological systems, where individuals increase their motility in response to local crowding and resource depletion, giving rise to self-organized clusters that resist coarsening\@~\cite{Koppel.2008,Liu.2013}.
This suggests that differential mobility may represent a universal strategy for maintaining spatial structure in out-of-equilibrium systems\@~\cite{Grosberg.2015,Weber.2016}.

\begin{acknowledgments} 
    We thank Henrik Weyer and Tobias Roth for stimulating discussions and Jan Willeke and Xaver Kainz for their insightful comments on our manuscript.
    This research was funded by the Deutsche Forschungsgemeinschaft (DFG, German Research Foundation) under Germany's Excellence Strategy – EXC3092 – 533751719.
    We further acknowledge support from the John Templeton Foundation, and the U.S. National Science Foundation (NSF) under Grant No.\@~PHY-2309135 through the Kavli Institute for Theoretical Physics (KITP).
\end{acknowledgments}


%

\end{document}